\documentstyle[aps,twocolumn,epsf]{revtex}  
\begin{document}
\twocolumn[\hsize\textwidth\columnwidth\hsize\csname 
@twocolumnfalse\endcsname                            
\title{Liquid-Gas phase transition in Bose-Einstein Condensates }
\author{A. Gammal$^{(a)}$, T. Frederico$^{(b)}$, L. Tomio$^{(a)}$ 
and Ph. Chomaz$^{(c)}$}
\address{$^{(a)}$ Inst. de F\'{\i}sica Te\'{o}rica, 
Universidade Estadual
Paulista, \\
01405-900 S\~{a}o Paulo, Brazil \\
$^{(b)}$Dept. de F\'{\i }sica, Instituto Tecnol\'{o}gico da
Aeron\'{a}utica, \\
Centro T\'{e}cnico Aeroespacial, 12228-900 S\~{a}o Jos\'{e} dos Campos, SP,
Brazil \\
$^{(c)}$ GANIL, B.P. 5027, F-14021 Caen Cedex, France }
\maketitle
\begin{abstract}
We study the effects of a repulsive three-body interaction on a system of
trapped ultra-cold atoms in a Bose-Einstein condensed state. The
corresponding $s-$wave non-linear Schr\"{o}dinger equation is solved
numerically and also by a variational approach. A first-order liquid-gas
phase transition is observed for the condensed state up to a critical
strength of the effective three-body force. 
\end{abstract}
\vskip 0.5cm ]                              

The experimental evidences of Bose-Einstein condensation (BEC) in
magnetically trapped weakly interacting atoms~\cite{and96,mew96,brad97}
brought a considerable support to the theoretical research on bosonic
condensation. The nature of the effective atom-atom interaction determines
the stability of the condensed state: the two-body pseudopotential is
repulsive for a positive $s-$wave atom-atom scattering length and it is
attractive for a negative scattering length~\cite{huang}. The ultra-cold
trapped atoms with repulsive two-body interaction undergoes a Bose-Einstein
phase-transition to a stable condensed state, in a number of cases found
experimentally, as for $^{87}$Rb~\cite{and96}, for $^{23}$Na~\cite{mew96}
and $^{7}$Li~\cite{brad97}. \ However, a condensed state of atoms with
negative $s-$wave atom-atom scattering length would be unstable for a large
number of atoms \cite{rup95,baym96}.

It was indeed observed in the $^{7}$Li gas~\cite{brad97}, for which the 
$s-$wave scattering length is $a=(-14.5\pm 0.4)$ \AA , that the number of
allowed atoms in the condensed state was limited to a maximum value between
650 and 1300, which is consistent with the mean-field  prediction~\cite
{rup95}. An earlier experiment~\cite{brad95} suggested that the number of
atoms in the condensate state was significantly larger than the theoretical
predictions with two-body pseudopotential. This is consistent with an
addition of a repulsive three-body interaction, which can extend
considerably the region of stability for the condensate even for a very weak
three-body force. \ 

It was reported in Ref.~\cite{esry} that a sufficiently dilute and cold
bosonic gas exhibits similar three-body dynamics for both signs of the 
$s-$wave atom-atom scattering length and the long-range three-body interaction
between neutral atoms is effectively repulsive for either sign of the
scattering length. 
It was suggested that, for a large number of bosons the three-body repulsion
can overcome the two-body attraction, and a stable condensate will appear in
the trap~\cite{josse}.
Singh and Rokhsar\cite{SinghRokhsar} have also observed that above the
critical value $n$ (which is proportional to their $-\gamma_c$) the only local
minimum is a dense gas state, where the neglect of three-body collisions fails. 

In this work, using the mean-field approximation, we investigate the
competition between the leading term of an attractive two-body interaction,
which is originated from a negative two-atom $s-$wave scattering length, and
a repulsive three-body interaction, which can happen in the 
Efimov limit~\cite{efimov} ($|a|\rightarrow \infty $) as discussed in 
Ref.~\cite{esry}~\footnote
{The physics of three-atoms in the Efimov limit 
is discussed in Ref.~\cite{3atom}. This reference extends a previous study 
of universal aspects of the Efimov effect~\cite{halo}. 
The relevance of three-body effects in BEC was also previously reported in 
Refs.~\cite{ADV,GFT}, where it is discussed the stability of 
the numerical solutions.}
We show that a kind of liquid-gas phase transition appears inside the Bose 
condensate.

The Ginzburg - Pitaevskii - Gross (GPG) non\-linear Schr\"{o}\-dinger equation
(NLSE) \cite{gin} is extended to include the effective potential coming 
from the three body interaction and then solved numerically in the
$s-$wave channel. 
The dimensionless parameters are related to the
two-body scattering length, the strength of the three-body interaction
and the number of atoms in the condensed state. 
As particularly observed in Ref.~\cite{hs}, to incorporate all 
two-body scattering processes in such many particle system, 
the two-body potential should be replaced by the many-body 
$T-$matrix. 
\ \ Usually, at very low energies, this is approximated by 
the two-body scattering matrix, which is directly proportional to the 
scattering length $a$~\cite{baym96}. 
To obtain the desired equation, we first consider the effective Lagrangian
density, which describes the condensed wave-function in the Hartree
approximation, implying in the GPG energy functional~\cite{gin} for the
trial wave function 
$\Psi $: 
\begin{eqnarray}
{\cal {L}}&=&\frac{i\hbar }{2}\left( \Psi ^{\dagger }\frac{
\partial \Psi }{\partial t}-\frac{\partial \Psi ^{\dagger }}{\partial t}\Psi
\right) +\frac{\hbar ^{2}}{2m}\Psi ^{\dagger }\nabla ^{2}\Psi
\nonumber\\ && 
- \frac{m}{2}\omega ^{2}r^{2}|\Psi |^{2}+ {\cal {L}}_{{\rm
{I}}}\ .  
\label{lag} 
\end{eqnarray}
In our description, the atomic trap is given by a rotationally symmetric
harmonic potential, with angular frequency $\omega $, and 
${\cal {L}}_{{\rm {I}}}$ gives the effective atom interactions up to three 
particles.

The effective interaction Lagrangian density for ultra-low temperature
bosonic atoms, including two~\cite{baym96} and three-body effective
interaction at zero energy, is written as: 
\begin{equation}
{\cal L}_{{\rm I}}=-\frac{2\pi \hbar ^{2}a}{m}\left| \Psi \right|
^{4}-\frac{2\lambda _{3}}{3{\rm {!}}}\left| \Psi \right| ^{6}\ ,
\label{lagI}
\end{equation}
where $\lambda _{3}$ is the strength of the three-body effective interaction
and $a$ the scattering length.

The NLSE, which describes the condensed wave-function in the mean-field
approximation, is variationally obtained from the effective Lagrangian given
in Eq.~(\ref{lag}). By considering 
a stationary solution, $\Psi (\vec{r},t)=e^{-i\mu t/\hbar }$ $\psi (\vec{r})$
where $\mu $ is the chemical potential and $\psi (\vec{r})$ is normalized to
1 and by rescaling the NLSE for the $s-$wave solution, we obtain 
\begin{equation}
\left[ -\frac{d^{2}}{dx^{2}}+\frac{1}{4}x^{2}-\frac{|\Phi (x)|^{2}}{x^{2}}
+g_{3}\frac{|\Phi (x)|^{4}}{x^{4}}\right] \Phi (x)\ =\ \beta \Phi (x)\ 
\label{schd}
\end{equation}
for $a<0$, where $x\equiv \sqrt{{2m\omega }/{\hbar }}\ r$ and $\Phi (x)\equiv
N^{1/2}\sqrt{8\pi |a|}\;r\psi (\vec{r})$. 
The dimensionless parameters, related to
the chemical potential and the three-body strength are, respectively, given
by $\beta \equiv \mu /\hbar \omega $ and $g_{3}\equiv \lambda _{3}\hbar
\omega m^{2}/(4\pi \hbar ^{2}a)^{2}$. The normalization for $\Phi (x)$ reads 
$\int_{0}^{\infty }dx|\Phi (x)|^{2}\ =n$ where the reduced number $n$ is
related to the number of atoms $N$ by $n\equiv 2N|a|\sqrt{2m\omega /\hbar }.$
The boundary conditions ~\cite{rup95} in Eq.(\ref{schd}) are given by
$\Phi(0)=0$ and $\Phi (x)$ $\to C\exp (-x^{2}/4+[\beta -\frac{1}{2}]\ln (x))$
when $x\to \infty .$ 

The above equation, (\ref{schd})  will be treated by numerical procedures
for non-linear differential equations, employing the Runge-Kutta 
(RK) and shooting methods.
However, it will be helpful first to consider a variational 
procedure~\cite{fetter}, using a trial Gaussian wave-function for 
$\psi (\vec{r})$
\begin{equation}
\psi _{var}(\vec{r})=\left( \frac{1}{\pi \alpha ^{2}}\frac{m\omega }{\hbar }
\right) ^{\frac{3}{4}}\exp {\left[ -\frac{r^{2}}{2\alpha ^{2}}\left( \frac{
m\omega }{\hbar }\right) \right] },  \label{varwf}
\end{equation}
where $\alpha $ is a dimensionless variational parameter. The corresponding
root-\-mean-\-square radius is proportional to the variational parameter 
$\alpha $, as $\left\langle r^{2}\right\rangle_{var}$ $=\alpha ^{2}3\hbar
/(2m\omega) $, 
while the central density is given by $\rho _{c,var}(\alpha )$ 
$=\alpha^{-3}\left(m\omega /\pi \hbar \right) ^{3/2}$. 
The expression for the total variational
energy is given by 
\begin{equation}
E_{var}(\alpha )=\hbar \omega N\left[ \frac{3}{4}\left( \alpha ^{2}+\frac{1}{
\alpha ^{2}}\right) -\frac{n}{4\sqrt{\pi }\alpha ^{3}}+\frac{2n^{2}g_{3}}{9
\sqrt{3}\pi \alpha ^{6}}\right] .  \label{Evar}
\end{equation}
In the same way, we can obtain the corresponding chemical potential , Eq.~(
\ref{schd}): 
\begin{equation}
\mu _{var}(\alpha )=\hbar \omega \left[ \frac{3}{4}\left( \alpha ^{2}+\frac{1
}{\alpha ^{2}}\right) -\frac{n}{2\sqrt{\pi }\alpha ^{3}}+\frac{2n^{2}g_{3}}{3
\sqrt{3}\pi \alpha ^{6}}\right] .  \label{muvar}
\end{equation}
The variational solutions of $E_{var}(\alpha )$ are given, as a function of $
n$ and $g_{3}$ (where $a<0$ and $g_{3}>0$), by finding the extrema of 
Eq.~(\ref{Evar}) with respect to $\alpha$.

\begin{figure}
\setlength{\epsfxsize}{1.0\hsize} \centerline{\epsfbox{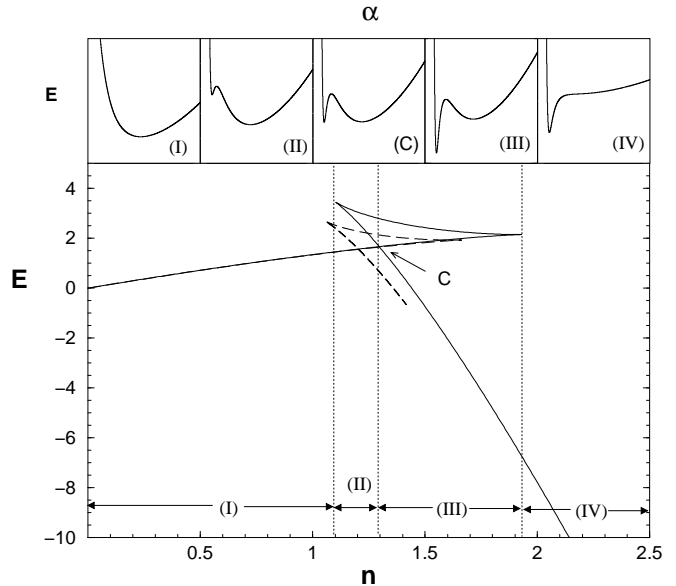}}

{\caption{In the lower part, we have a comparison between variational 
(solid curve)
and exact (dashed curve) numerical calculations of the condensate energy as
a function of the reduced number of atoms $n$ for $g_{3}=0.005\ .$ In the
upper frame we show five  plots of the variational energy as a function of
the variational parameter $\alpha $ for  five particular values of $n$ shown
also in the lower frame. (I) (resp IV) corresponds to a small (large)  $n$
region where only one stable solution is encountered; (II) (resp III) to a
small (large) $n$ region where we observe three extrema for the energy;
(C) corresponds to a particular $n$ where we obtain two stable solutions
with the same energy $E_{1}=E_{2}$.   $E$ is given in units of $(N\hbar
\omega )/n$.}}
\end{figure}
In Fig. 1, we first illustrate the variational procedure 
considering an arbitrarily small three-body interaction, chosen as $
g_{3}=0.005$. In the upper part of the figure, we show five small plots for
the total variational energy $E$, in terms of the variational width $\alpha $. 
Each one of the small plots corresponds to particular values of $n$. 
For each number $n$ we  report the energy of the variational extrema in
the lower part of figure 1. 
In region (I) where the number of atoms is still small, the attractive two
body force dominates over the repulsive three-body force and just one minima
of the energy as a function of the variational parameter $\alpha $ is found. 
That is also the case for $g_{3}=0$. 
When the number of atoms is further increased (region (II)) two minima
appear in the energy $E\left( \alpha \right) .$ An unstable maximum
is also found between the two minima. The lower energy minimum is stable while the
solution corresponding to the smaller $\alpha $ is metastable. 
This solution has a higher density  
and, consequently, its metastability is justified by the repulsive
three-body force acting at higher densities. The minimum number $n$ for the
appearance of the metastable state is characterized by an inflection point
in the energy as a function of $\alpha $. The value of $n$ at the inflection
point corresponds to the beak in the plot of extremum energy versus $n$
because for larger $n$ three variational solutions are found as depicted in
the lower part of figure 1. The attractive two-body and trap potentials
dominate
the condensed state in the low-density stable phase up to the crossing point
(C). At this point, the denser metastable solution becomes degenerate in
energy with the lower-density stable solution and a first order phase
transition takes place. Since the two solutions differ by their density this
transition is analogous to a gas-liquid phase transition for which the
density difference between the liquid and the gas is the order parameter. In
the variational calculation this occurs at the transition number 
$n\approx$1.3
while the numerical solution of the NLSE gives $1.2$.
In region (III), we observe two local minima with different energies, a
higher-density stable point and a lower-density metastable point. The
metastable solution disappears in the beak at the boundary between region
(III) and (IV). In regions (III) and (IV) the three-body repulsion
stabilized a dense solution against the collapse induced by the two-body
attraction. The qualitative features of the variational solution is clearly
verified by the numerical solution of the NLSE, as shown by the dashed curve.

\begin{figure}
\setlength{\epsfxsize}{1.0\hsize} \centerline{\epsfbox{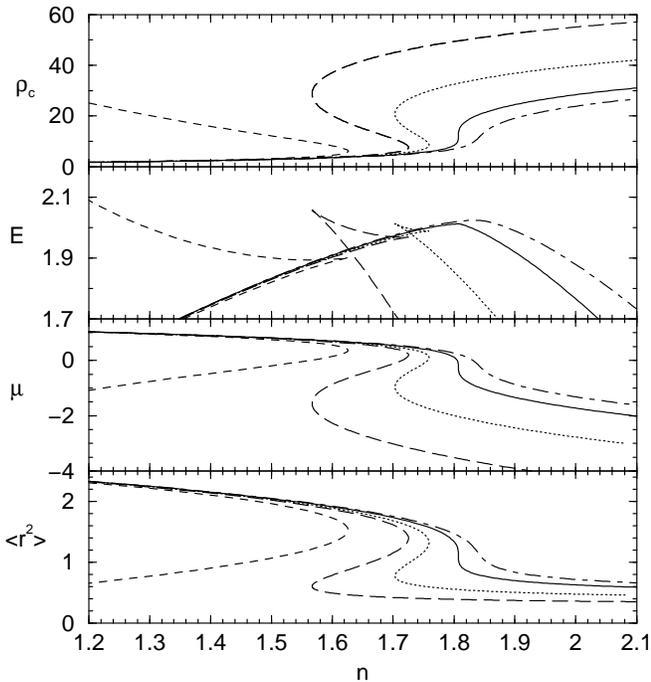}}

\caption{ Central density $\rho _{c}$, total energy $E$, chemical
potential $\mu $, and average square radius $\left\langle r^{2}\right\rangle 
$, as functions of the reduced number of atoms $n.$ Several values of the
dimensionless three body interaction strength $g_{3}$ were used : $g_{3}=0$
(short-dashed line), $g_{3}=0.012$ (dashed line), $g_{3}=0.015$ (dotted
line), $g_{3}=0.0183$ (solid line), $g_{3}=0.02$ (dot-dashed line). The
corresponding units are: $(m\omega /\hbar )/(4\pi |a|)$ for $\rho _{c}$, $
(N\hbar \omega )/n$ for $E$, $\hbar \omega $ for $\mu $, and $\hbar
/(2m\omega )$ for $\left\langle r^{2}\right\rangle $.}
\end{figure}
In Fig. 2, considering several values of $g_{3}$ (0, 0.012, 0.015, 0.0183
and 0.02), using exact numerical calculations, we present the evolution of some
relevant physical quantities $E,$ $\mu ,$ $\rho _{c}$ and $\ \left\langle
r^{2}\right\rangle $ as functions of the reduced number of atoms $n$.
\ For $g_{3}=0$, our calculation reproduces the result presented in 
Ref.~\cite{rup95,hs}, with the maximum number of atoms limited by 
$n_{max}\approx 1.62$ \ ($n$ is equal to $|C_{nl}^{3D}|$ of 
Ref.~\cite{rup95}). In the plot for
the energy as a function of $n$ it is shown 
that for values of $g_{3}\ >\ 0.0183$ the phase transition is absent. At $
g_{3}\approx 0.0183$ and $n\approx 1.8$, the stable, metastable and unstable
solutions come to be the same. This corresponds to a critical point
associated with a second order phase transition. At this point the
derivatives of $\mu ,$ $\rho _{c}$ and $\ \left\langle r^{2}\right\rangle $
as a function of $n$ all diverge.

As shown in the figure, for $0<\ g_{3}\ <\ 0.0183$, the density $\rho _{c},$
the chemical potential $\mu $ and the root-mean-squared radius $\left\langle
r^{2}\right\rangle $ present back bendings typical of a first order phase
transition. For each $g_{3},$ the transition point given by the crossing
point in the $E$ versus $n$ corresponds to a Maxwell construction in the
diagram of $\mu $ versus $n$. At this point an equilibrated
condensate should undergo a phase transition from the branch extending to
small $n$ to the branch extending to large $n.$ The system should never
explore the back bending part of the diagram because as we have seen in
figure 1 it is an unstable extremum of the energy. From this figure it is
clear that the first branch is associated with large radii, small densities
and positive chemical potentials while the second branch presents a more
compact configuration with a smaller radius a larger density and a negative
chemical potential. This justify the term gas for the first one and liquid
for the second one. However we want to stress that both solutions are
quantum fluids. With $g_{3}=0.012$ the gas phase happens for $n<1.64$ and
the liquid phase for $n>1.64$. For $g_{3}\ >\ 0.0183$ all the presented
curves are well behaved and a single fluid phase is observed.
We also checked that calculations with the variational expression of
$\langle r^2\rangle$, $\rho_c$ and $\mu$ are in good agreement with the
ones depicted in Fig.2, following the same trend shown in Fig.1 for the 
energy.

Finally, in the lower frame of Fig. 3, we show the phase boundary separating 
the two phases in the plane defined by $n$ and $g_{3}$ and the critical point 
at $n\approx 1.8$ and $g_{3}\approx 0.0183$. In the upper frame, we show the
boundary of the forbidden region in the central density versus  
$g_{3}$ diagram. 

To summarize, our calculation presents, at the mean-field level, the
consequences of a repulsive three-body effective interaction for the Bose
condensed wave-function, together with an attractive two-body interaction. A
first-order liquid-gas phase-transition is observed for the condensed state
as soon as a small repulsive effective three-body force is introduced. In 
dimensionless units the critical point is obtained when $g_{3}\approx 0.0183$
and $n\approx 1.8$. The characterization of the two-phases through their
energies, chemical potentials, central densities and radius were also given
for several values of the three-body parameter $g_{3}$. 

\begin{figure}
\setlength{\epsfxsize}{1.0\hsize} \centerline{\epsfbox{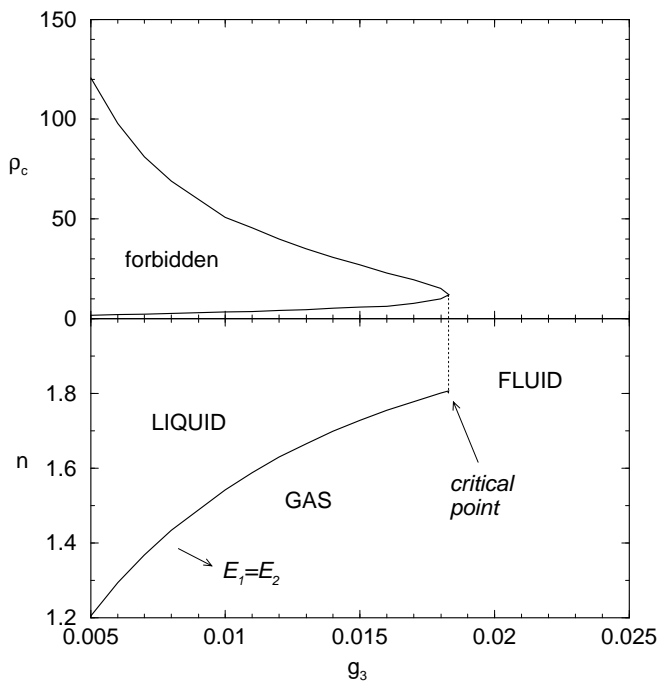}}

\caption{Phase diagram of the Bose condensate. Central density
$\rho_c $ \ in units of $(m\omega/\hbar)/(4\pi |a|).$}
\end{figure}
The results presented in this paper can be relevant to determine a possible
clear signature of the presence of repulsive three-body interactions in
Bose condensed atoms. It points to a new type of phase transition between
two Bose fluids. Because of the condensation of the atoms in a single
wave-function this transition may present very peculiar fluctuations and
correlations properties. As a consequence, it may fall into a different
universality class than the standard liquid-gas phase transition, which are
strongly affected by many-body correlations. This question certainly
deserves further studies.

{\bf Acknowledgments}

We thank the organizers of the ``International Workshop on Collective
Excitations in Fermi and Bose Systems" (S\~ao Paulo, Brazil), C. Bertulani,
L.F. Canto and M. Hussein, to provide the conditions for stimulating
discussions and for the starting of this collaboration. 
AG, TF and LT also thank N. Akhmediev and M.P. Das for correspondence
related to Ref.\cite{GFT}, after conclusion of this work,
which has pointed out their previous suggestion of relevance of three-body
effects in BEC~\cite{ADV}. \ 
This work was partially supported by Funda\c c\~ao de Amparo \`a Pesquisa
do Estado de S\~ao Paulo and Conselho Nacional de Desenvolvimento
Cient\'\i fico e Tecnol\'ogico.

\end{document}